\documentclass[aps,prb,twocolumn,superscriptaddress,floatfix]{revtex4}
\usepackage{graphicx}
\usepackage{dcolumn}
\usepackage{bm}
\usepackage{stmaryrd}
\usepackage{latexsym}
\usepackage{amssymb}
\usepackage{amsfonts}
\usepackage{amsmath}
\usepackage{fancybox}
\usepackage{color}
\usepackage{verbatim}
\def\kv{{\bm k}}
\def\kpv{{\bm k}'}

\def\be{\begin{equation}}
\def\ee{\end{equation}}
\def\ber{\begin{eqnarray}}
\def\eer{\end{eqnarray}}

\begin{document}
\title{Persistent Current States in Bilayer Graphene}
\author{Jeil Jung}
\email{jeil.jung@gmail.com}
\affiliation{Department of Physics, University of Texas at Austin, Austin, Texas 78712, USA}
\affiliation{Department of Physics, University of Seoul, Seoul, 130-743, Korea}
\author{Marco Polini}
\affiliation{NEST, Istituto Nanoscienze-CNR and Scuola Normale Superiore, I-56126 Pisa, Italy}
\author{A.H. MacDonald}
\affiliation{Department of Physics, University of Texas at Austin, Austin, Texas 78712, USA}

\begin{abstract}
We argue that at finite carrier density and large displacement fields,
bilayer graphene is prone to $\ell =0$ and $\ell = 1$ Pomeranchuk Fermi surface instabilities.  
The broken symmetries are driven by non-local exchange interactions 
which favor momentum space condensation.
We find that electron-electron interactions lead first to spontaneous valley polarization, 
which breaks time-reversal invariance and is associated with spontaneous orbital 
magnetism, and then under some circumstances to a nematic phase with 
reduced rotational symmetry.  When present, nematic order is signaled by 
reduced symmetry in the dependence of optical absorption on light polarization.  
\end{abstract}
%


\maketitle

\section{Introduction} 

The search for a microscopic theory of superconductivity, ultimately 
brought to a conclusion by the successful work 
of Bardeen, Cooper, and Schrieffer~\cite{bcstheory}, led a number of early 
condensed matter theory researchers~\cite{heisenberg,born,london}
to speculate on the possibility that interactions could under some circumstances 
lead to momentum-space order responsible for equilibrium currents.  
Since orbital magnetism always accompanies  spontaneous spin-polarization
because of spin-orbit coupling, we now know that 
$\ell=0$ Pomeranchuk~\cite{pomeranchuk_jetp_1958} instabilities can 
lead to spontaneous circulating currents.  Similarly $\ell=1$ Fermi surface 
instabilities can potentially lead to spontaneous longitudinal currents.  

\begin{figure}[t]
\label{schematic}
\includegraphics[width=7.5cm,angle=0]{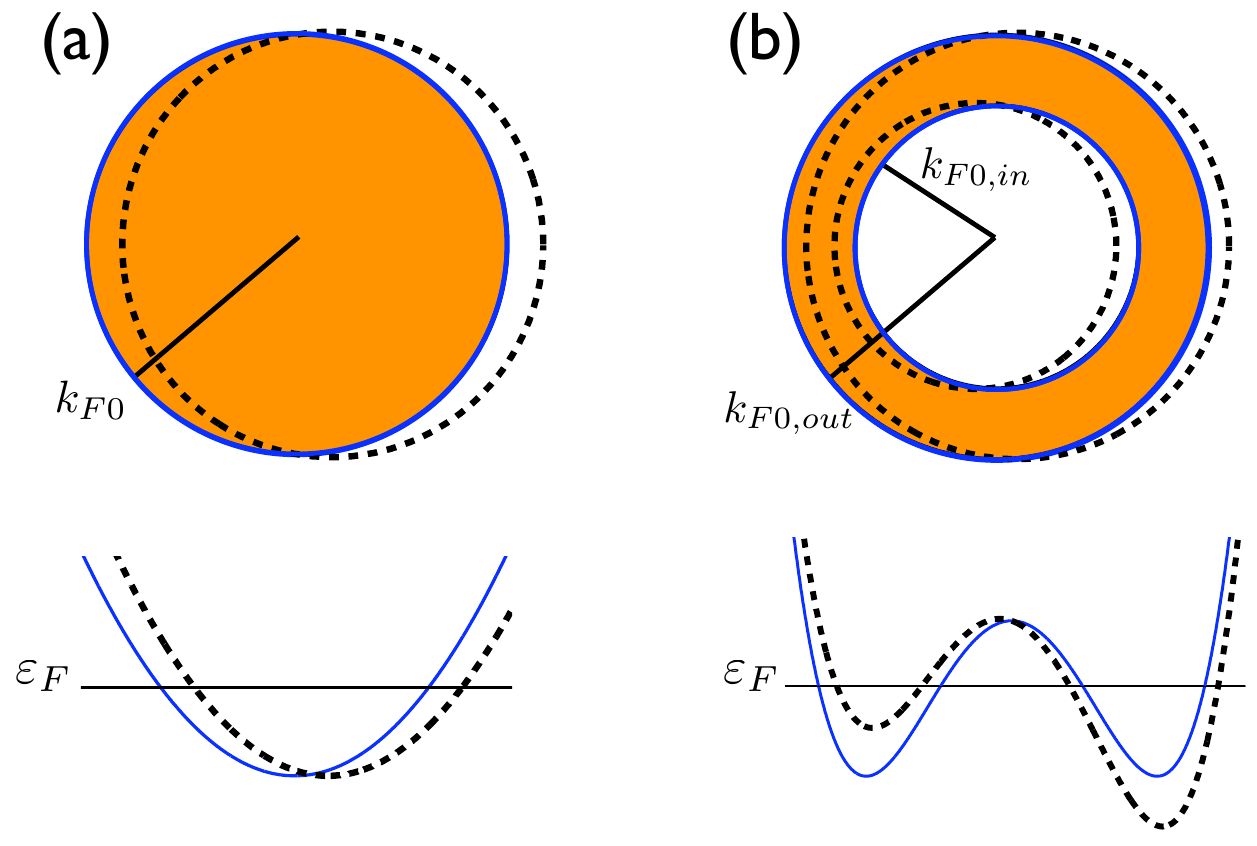} 
\caption{(color online) 
$\ell=1$ Pomeranchuk instability of a conventional Fermi liquid and of density-unbalanced BLG.  
The lower panels illustrate the quasiparticle band dispersions that
correspond to the Fermi surfaces illustrated in the 
top panels. 
a) The circular orange Fermi circle with Fermi radius $k_{{\rm F}0}$ denotes the occupied states of a two-dimensional Fermi sea.
A Pomeranchuk instability occurs when energy is reduced by a momentum-space angle-dependent change in Fermi radius
that is proportional to $\cos(\ell\theta)$.  For $\ell=1$ the distorted Fermi surface, indicated here by the dashed black circle, 
is simply shifted in momentum space.
b) The disk-shaped orange region with inner and outer Fermi radii
$k_{{\rm F}0, {\rm in}}$ and $k_{{\rm F}0, {\rm out}}$ denotes
occupied conduction band states in unbalanced BLG.  Shifts in the 
inner and outer Fermi radii that are proportional to $\cos(\theta)$ and have opposite 
signs for the inner and outer Fermi lines, thicken and thin the disk in opposite directions 
as indicated by the black dashed lines.  The distorted phase has lower exchange energy
because the occupied states are more concentrated in momentum space.\label{fig:one}}
\end{figure}

An $\ell=1$ Fermi-surface distortion transfers occupation between  
quasiparticle states with opposite current components along the 
continuously variable direction in which overall current flows, 
as illustrated schematically in Fig.~\ref{fig:one}.
For a translationally-invariant electronic system it is known  
from both microscopic-quantum-mechanical~\cite{bohm,smith} and Fermi-liquid-theory~\cite{Vignale} 
points of view that such a distortion
corresponds simply to a Galilean boost which raises the 
center-of-mass kinetic energy and does not change the interaction energy.
An $\ell=1$ charge-channel Pomeranchuk instability is therefore an 
impossibility in an electron fluid.  In this article we show how the simple, but 
highly unusual electronic structure of graphene bilayers with low carrier densities and 
large displacement fields can lead to both $\ell=0$ and $\ell=1$ Pomeranchuk instabilities,
although not to spontaneous longitudinal currents.  

In a crystal the Pomeranchuk instability notion 
refers to Fermi-surface distortions which reduce lattice symmetries.
Here $\ell=1$ charge-channel distortions have also been viewed~\cite{Schofield} as extremely unlikely,
although instabilities in $p$-wave spin channels preserving time reversal symmetry have been proposed
under certain conditions.~\cite{pwave}
Instabilities with $\ell=2$, which lead to electron nematic states, do on the 
other hand appear to occur~\cite{Metzner,fradkin1} in a variety of different systems
and spin-channel $\ell=0$ instabilities, which lead to ferromagnetism, are of course common.
In this article 
we point out that a Bernal-stacked graphene bilayer (BLG) in which a gap has been opened by a 
transverse electric field is ideally suited to host both $\ell=0$ and $\ell=1$ Pomeranchuk instabilities
because its electron and hole Fermi seas are disks that are spread widely over momentum space, 
as illustrated in Fig.~\ref{fig:one}(b).~\cite{pomeranchuk_single_layer}
Both $\ell=0$ and $\ell=1$ Fermi surface distortions can lower energy by compactifying the Fermi 
sea, thereby realizing the momentum-space condensation envisaged by London~\cite{london} 
more than seventy years ago.

\section{Fermi-Surface Instabilities in Bilayer Graphene} 

Recent progress~\cite{graphenereviews} has made it possible to prepare and 
study the electronic properties of two-dimensional (2D) electron systems based on 
single- and few-layer graphene.  This advance has
provided researchers with a new family of materials whose 
electronic structure is at the same time remarkably simple and remarkably variable.  
Single-layer graphene is described by a massless-Dirac-fermion model with 
conduction and valence bands that touch at two different points in momentum 
space and disperse linearly over a wide energy region.
The parabolic band dispersion near the Fermi level of a neutral Bernal-stacked bilayer graphene 
allows electron-interaction-driven instabilities, which have been studied by using mean-field~\cite{meanfield,hfbilayer} 
or renormalization-group-based approaches.~\cite{rg} 
Clear evidence for strong many-body effects has already been
obtained in several recent experiments.~\cite{velasco,geim,yacoby,freitag,pnas,paper1,paper2,paper3,paper4,paper5,paper6,paper7}.    
In a neutral Bernal-stacked bilayer with a very strong  
electric field directed perpendicular to the layers~\cite{Castro_review}, 
the Fermi level lies within the conduction band of one layer and the valence band of the 
other.  Inter-layer hybridization with strength $\gamma_1$ opens up an avoided crossing gap
centered at a finite 2D wavevector magnitude between states localized in opposite layers,
yielding an unusual semiconductor with an electrically tunable gap.  
Our interest here is in the electronic properties of degenerate electrons in the conduction band 
(or holes in the valence band)
of this layer-unbalanced configuration of BLG.  

When angular variation of the avoided 
crossing gap due to trigonal warping effects is neglected,~\cite{trigonalwarping} the conduction
band minimum occurs along a circle in momentum space and the non-interacting 
electron Fermi surface is an annulus as indicated in Fig.~\ref{fig:one}(b). 
Because the band density-of-states diverges as energy approaches the 
conduction band minimum it is clear, as noted previously by others,~\cite{CastroNeto,CDW}
that interactions may play a central role in determining electronic properties 
and that instabilities that break symmetries are likely. 
The possibilities include ferromagnetism~\cite{CastroNeto} and  
density-wave~\cite{CDW} states.  In this paper we argue that momentum space 
condensation of the type imagined by London, Heisenberg and others, which has not previously been 
observed, is also a possibility.

\begin{figure}[t!]
\includegraphics[width=1\linewidth]{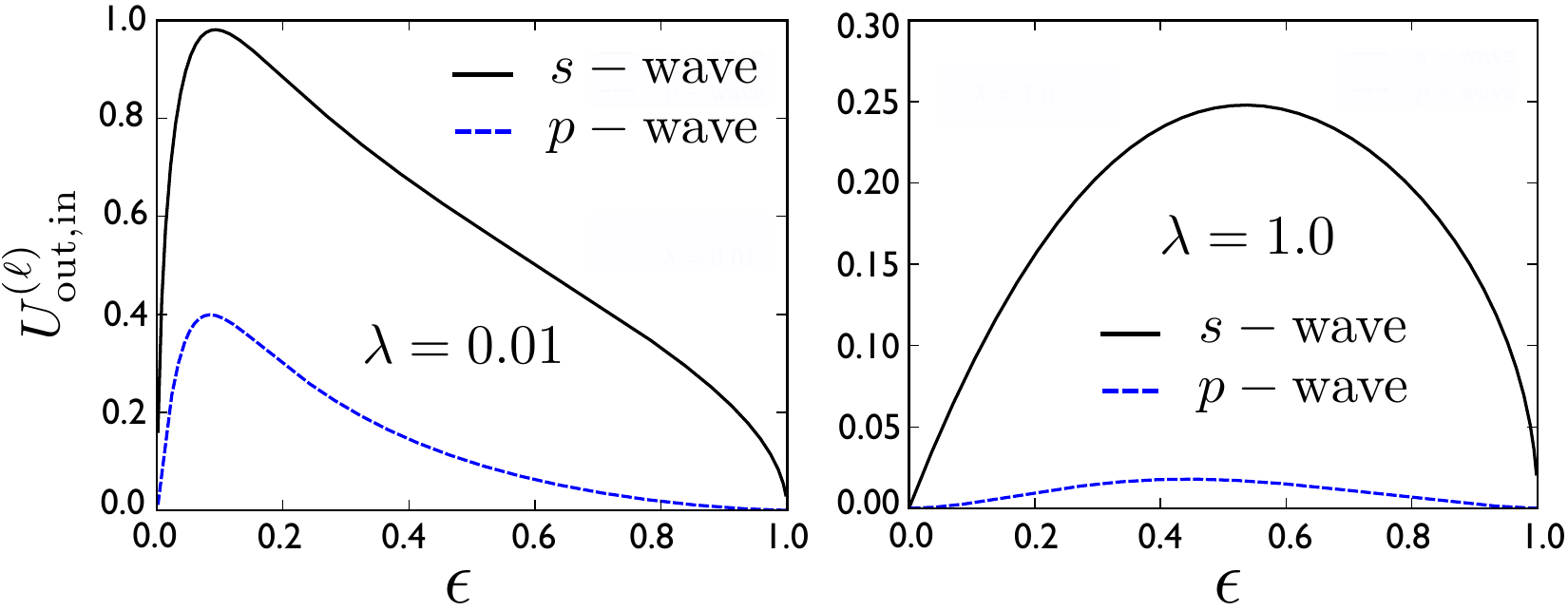} 
\caption{(color online) 
The dimensionless ``outer-inner" pseudopotential $U^{(\ell)}_{{\rm out}, {\rm in}}$ {\em vs.} $\epsilon = b/(2{\bar k})$ 
for $\ell =0$ ($s$-wave) and $\ell =1$ ($p$-wave).  This plot is for $\delta  = 2/(a_{\rm eff} {\bar k}) = 1$
where $a_{\rm eff}$ and ${\bar k}$ are defined in the main text. 
The parameter $b$ measures the thickness of the ring
and $b \to 0$ in the limit in which the Fermi energy $\varepsilon_{\rm F}$ equals the band-edge energy 
$\epsilon_{\rm min}$. The left panel is for a very small screening parameter, $\lambda =0.01$, 
while the right panel is for fully screened Thomas-Fermi interactions, 
$\lambda=1.0$.\label{fig:two}}
\end{figure}

We illustrate our main point by considering a toy model which ignores
trigonal warping and spin and valley degrees-of-freedom, and by 
using mean-field theory to estimate its Fermi-liquid parameters.   
The Hartree-Fock energy functional~\cite{hartreecaveat} is
\be\label{eq:HFfreeenergy}
E_{\rm HF}[\{n_\kv\}] = \sum_{\kv} [\varepsilon_{\rm b}(k) - \mu]n_\kv -\frac{1}{2A}\sum_{\kv, \kpv} n_\kv V_{\kv -\kpv} n_{\kpv}~,
\ee
where $\varepsilon_{\rm b}(k)$ are bare-band energies, which are isotropic and thus 
depend only on $k =|\kv|$, and $A$ is the 2D electron system area. 
In our toy model we assume that the interaction $V_{\kv -\kpv}$ depends only on $|\kv -\kpv|$. 
Expanding this energy functional in powers of the deviation $\delta n_\kv$ from the occupation numbers $n^{(0)}_{\kv}$ corresponding to the 
undistorted Fermi surface yields an energy expression of the Fermi-liquid-theory form: 
\be \label{eq:HFfunctional}
E_{\rm HF}[\{n_\kv\}] = E_0 + \sum_{\kv}\varepsilon^{\rm HF}_{\rm b}(k) \delta n_\kv - \frac{1}{2A}\sum_{\kv, \kpv} \delta n_\kv V_{\kv -\kpv} \delta n_{\kpv}~,
\ee
where the Hartree-Fock band energy $\varepsilon^{\rm HF}_{\rm b}(k)$ is defined by
\ber\label{eq:HFvelocity}
\varepsilon^{\rm HF}_{\rm b}(k) = \varepsilon_{\rm b}(k) - \int \frac{d^2 \kpv}{(2\pi)^2} n^{(0)}_{\kpv} V_{\kv -\kpv}~.
\eer
Because the total energy is prone to cancellations between quasiparticle-velocity renormalizations
and quasiparticle-interaction effects, we must treat the two contributions on an equal footing. 
Below we consider the quasiparticle-velocity renormalization first and use this analysis to define our notation.

\begin{figure*}[t!]
\includegraphics[width=15cm,angle=0]{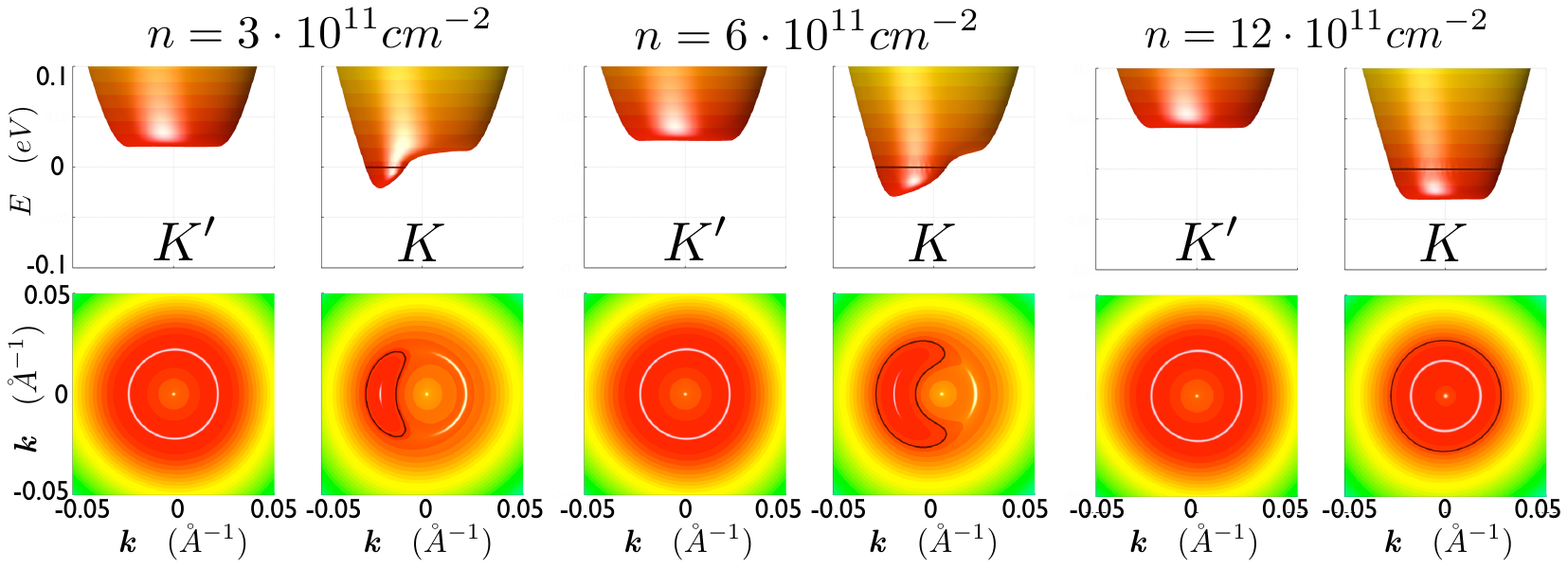}
\caption{(color online) 
Valley polarization and deformation of quasiparticle bands
due to $\ell = 0$ and $\ell = 1$ Pomeranchuk instabilities
in unbalanced BLG.  
These results were obtained with an external potential difference 
corresponding to a perpendicular electric field of $1~{\rm V}/{\rm nm}$ 
between layers.  
Valley polarized quasiparticle bands are plotted for three different total carrier densities.
The upper and lower panels illustrate conduction band energy dispersions  
and constant energy contours respectively. The left and right columns are for
momenta near the $K'$ and $K$ valleys. The $\ell = 0$ instability 
is signaled by unequal occupation of the two valleys. 
The $\ell = 1$ instability is signaled by broken rotational symmetry within an occupied valley. 
The solid black line marks the position of the Fermi level, chosen as the zero
of energy: $\varepsilon_{\rm F} = 0$.
For this calculation we used a $k$-point sampling density near the Dirac cones equivalent to 
a density of $4608 \times 4608$ points in the whole Brillouin zone.
\label{fig:three}
}
\end{figure*}

We begin by calculating the Hartree-Fock velocities, {\it i.e.} the momentum-space 
radial derivatives of the quasiparticle energy at the outer ($n ={\rm out}$) and inner
($n ={\rm in}$) Fermi circles.
The velocities are most conveniently evaluated by replacing the derivative with respect to 
$\kv$ in $V_{\kv -\kpv}$ by a derivative with respect to $\kpv$.
When the interaction correction to the velocity 
is integrated by parts, this derivative then acts on $n^{(0)}_{\kpv}$ picking out 
states at the Fermi energy.  We find that 
\ber\label{eq:HFclear}
v^{\rm HF}_n(k) &=& {\rm sgn}(n)|v_n(k)| + \frac{1}{2\pi\hbar} \int_0^{2\pi} \frac{d\varphi}{2\pi}\cos(\varphi) \nonumber\\
&\times&\left\{[k'V_{\kv -\kpv}]_{k' = k_{{\rm F}0, {\rm out}}} - [k'V_{\kv -\kpv}]_{k' = k_{{\rm F}0, {\rm in}}}\right\}~,\nonumber\\
\eer
where $\phi$ is the difference between the angular coordinates of $\kv$ and $\kpv$,
and we have noted that $V_{{\bm k}-{\bm k}'}$ depends only on $\phi$ 
once $k$ and $k'$ are fixed \cite{borghi_ssc_2009}.  This leads to
\be\label{eq:HFout}
\frac{v^\star_{\rm out}}{v_{\rm out}} = 1 + \alpha_{\rm out} \left(U^{(1)}_{{\rm out}, {\rm out}} 
- \sqrt{\frac{k_{{\rm F}0, {\rm in}}}{k_{{\rm F}0, {\rm out}}}}U^{(1)}_{{\rm out}, {\rm in}}\right)~,
\ee
where $v^\star_{\rm out} \equiv v^{\rm HF}_{\rm out}(k_{{\rm F}0,{\rm out}})$, 
$\alpha_{\rm out} \equiv e^2/(\hbar v_{\rm out})$, 
$U^{(m)}_{n, n'}  \equiv V_m(k_{{\rm F}0, n},k_{{\rm F}0, n'})  \left( k_{{\rm F}0, n}\,k_{{\rm F}0, n'} \right)^{1/2}/(2\pi e^2)$ 
is a dimensionless interaction parameter that is symmetric in the $n,n'$ inner/outer indices, and the 
$V_m$'s are Fourier components of the interaction's $\phi$-dependence. 
The corresponding expression for  $v^\star_{\rm in}$ can be obtained by interchanging the ${\rm in}$ and 
$\rm{out}$ labels.

To look for Pomeranchuk instabilities we paramaterize the inner ($n ={\rm in}$) and outer ($n={\rm out}$) Fermi surfaces in terms of dimensionless distortion functions: 
\be\label{eq:distortion}
k_{{\rm F}, n} = k_{{\rm F}0, n}[1+ a_n(\theta)] \equiv k_{{\rm F}0, n} + \delta k_{{\rm F}, n}(\theta)~,
\ee
and expand these in terms of their angular momentum components 
$a_n(\theta) = \sum_{\ell= -\infty}^{+\infty} a_{n\ell} e^{i \ell \theta}$. 
The distorted state has a $\delta n_\kv$ which is non-zero only in the vicinity of the inner and outer Fermi lines. 
It is easy to verify that the first-order correction to the energy vanishes.  To obtain the second-order correction we linearize the Hartree-Fock energies around $k_{{\rm F}0, {\rm in}}$ and $k_{{\rm F}0, {\rm out}}$:
\be
\varepsilon^{\rm HF}_{\rm b}(k) \simeq 
\left\{
\begin{array}{l}
- \hbar v^\star_{\rm in} (k - k_{{\rm F}0, {\rm in}}) \vspace{0.1 cm}\nonumber\\
\hbar v^\star_{\rm out} (k - k_{{\rm F}0, {\rm out}})
\end{array}
\right.
\ee
and add the quasiparticle interaction contribution to obtain the energy change for small distortions: 
\ber
\frac{E^{(2)}}{A} &=& \frac{\hbar}{4\pi}
\sum_{n, n'}\sum_{\ell=-\infty}^{+\infty} \Big\{ v^\star_n \delta_{n, n'} \nonumber\\
&-&\frac{e^2}{\hbar}{\rm sgn}(nn') U^{(\ell)}_{n, n'}\Big\}~k^{3/2}_{{\rm F}0, n},k^{3/2}_{{\rm F}0, n'}~a_{n\ell}a^*_{n'\ell}~.\nonumber\\
\eer
This is our principal result.  Note that for a Galilean-invariant system which has only an outer 
Fermi radius the two interaction contributions to the $\ell=1$ distortion energy cancel,
recovering the {\em no go} theorem~\cite{bohm,smith,Vignale} mentioned previously.  
An $\ell =1$ Pomeranchuk instability occurs when the determinant of the matrix
\be {\bm F} =  
\left(
\begin{array}{cc}
v^\star_{\rm out} - (e^2/\hbar)U^{(1)}_{{\rm out}, {\rm out}} & (e^2/\hbar)U^{(1)}_{{\rm out}, {\rm in}}\\
(e^2/\hbar)U^{(1)}_{{\rm out}, {\rm in}} & v^\star_{\rm in} - (e^2/\hbar)U^{(1)}_{{\rm in}, {\rm in}}
\end{array}
\right)
\label{quadraticform} 
\ee
is zero. Using Eq.~(\ref{eq:HFout}) (and the corresponding equation for $v^\star_{\rm in}$) we finally find the following criterion:
\be\label{eq:finalcriterion}
1 - \left(\alpha_{\rm out}\sqrt{\frac{k_{{\rm F}0, {\rm in}}}{k_{{\rm F}0, {\rm out}}}} + 
\alpha_{\rm in}\sqrt{\frac{k_{{\rm F}0, {\rm out}}}{k_{{\rm F}0, {\rm in}}}}\right) U^{(1)}_{{\rm out}, {\rm in}} =0~.
\ee
Notice that this instability criterion depends only on the ``outer-inner" interaction.

\begin{figure*}[t!]
\includegraphics[width=18cm,angle=0]{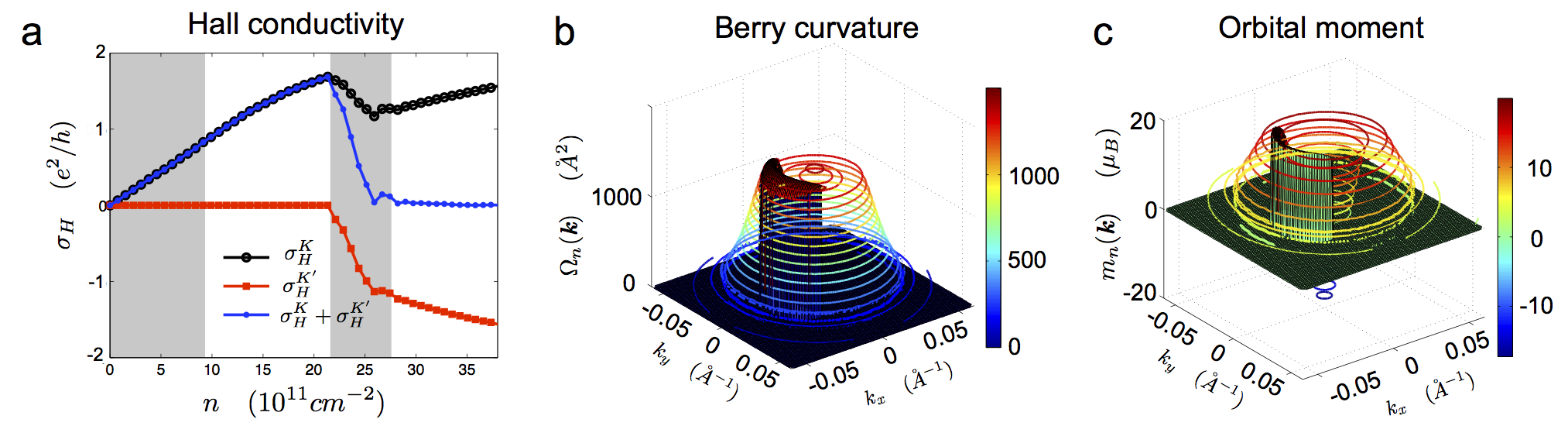}
\caption{(color online) 
{\bf a.} Spontaneous Hall effect {\em vs.} carrier density.  The Hall effect [Eq.~(\ref{hallcond})] 
reflects spontaneous valley polarization.
We find that the Hall conductivity increases monotonically with carrier density 
until the minority ($K'$) valley with opposite Hall conductivity starts to be occupied.
The shaded regions indicate intervals of carrier density over which 
nematic phases due to $\ell = 1$ and higher order channel Pomeranchuk instabilities appear.
The lower density instability occurs within the majority valley and the higher density 
instability within the minority valley.
{\bf b-c.} Berry curvature and Bloch state orbital magnetic moment in units of Bohr magnetons 
near the $K$ valley in the limit of low carrier density.
\label{fig:four}
}
\end{figure*}

We estimate the disk Fermi-liquid interaction parameters using 
\be \label{eq:screenedcoulomb}
V_{\kv -\kpv} = \frac{2\pi e^2}{|\kv - \kpv| +\lambda q_{\rm TF}}~,
\ee
where $q_{\rm TF}$ is the Thomas-Fermi screening wave vector,
and $\lambda \in [0,1]$ is a dimensionless control parameter that allows us to interpolate between bare-Coulomb ($\lambda =0$) 
and Thomas-Fermi ($\lambda=1$) limits. The Thomas-Fermi screening wave vector is proportional to the density-of-states at the Fermi energy, 
$N(0) = 2m_{\rm eff} {\bar k}/(\pi \hbar^2 b)$ where $b = k_{{\rm F}0, {\rm out}} - k_{{\rm F}0, {\rm in}}$, 
${\bar k}=  (k_{{\rm F}0, {\rm out}} + k_{{\rm F}0, {\rm in}})/2$, and $m_{\rm eff}$ parameterize the 
band-energy dispersion at its minimum.  It follows that $q_{\rm TF} = 2\pi e^2 N(0) = 4 {\bar k} / (a_{\rm eff} b)$ 
where $a_{\rm eff} = \hbar^2/(m_{\rm eff} e^2)$ is an effective Bohr radius. 
The inner-outer interaction parameters $U^{(\ell)}_{{\rm out}, {\rm in}}$ obtained using this 
approximation depend on two dimensionless quantities, $\epsilon = b/(2{\bar k})$ and $\delta = 2/(a_{\rm eff} {\bar k})$, and are plotted for $\ell = 0,1$ in Fig.~\ref{fig:two}. The rapid variation at small 
$\epsilon$ is due to strong Thomas-Fermi screening in this limit 
and are certainly an artifact of using such a simple screening approximation to
describe the $\epsilon \to 0$ limit in which Fermi-level density-of-states diverges. 
 
We can make two conclusions based on these estimates: i) Interactions in the $\ell=1$ channel are 
strong enough to produce an $\ell=1$ Pomeranchuk instability and ii) Interactions in the 
$\ell=0$ channel are almost certainly stronger than those in the $\ell=1$ channel.
A single valley, $\ell=0$ charge channel instability, would imply
phase separation into low and high density regions, 
is forbidden by the long-range Coulomb interaction.
However, the presence of valley and spin degrees of freedom allows 
$\ell=0$ instabilities that 
lead to spin~\cite{CastroNeto} or valley polarized states.  
$\ell=1$ Pomeranchuk instabilities
are likely only in states which are already spin or valley polarized.
Although momentum space is occupied asymmetrically when the $\ell=1$ instability occurs, 
we find that there is no longitudinal current because of a cancellation between inner and outer Fermi surface contributions.

\section{Tight-binding calculations and Hall conductivity signatures of the instabilities}

We have corroborated the conclusions made on the basis of our toy-model calculations 
by performing self-consistent $\pi$-band 
lattice Hartree-Fock calculations similar to those described in Refs.~\onlinecite{hfbilayer,hfmonolayer}. 
For these calculations we added interactions to a band   
model with nearest-neighbor intra-layer hopping $\gamma_0 = 3.12~{\rm eV}$ 
and inter-layer hopping  $\gamma_1 = 0.377~{\rm eV}$, ignoring other hopping
parameters for the sake of simplicity.~\cite{footnote}
The resulting bands near the $K$ and $K'$ points are shown in Fig. \ref{fig:three}.
We have suppressed spin-polarization instabilities in these calculations so that 
$\ell=0$ instabilities are manifested by spontaneous valley and not spin polarization.  
Because external electric fields in unbalanced BLG 
lead to large Berry curvatures $\Omega_{n, {\bm k}} $ of opposite sign in the vicinity 
of $K$ and $K'$ valley points~\cite{xiao} the anomalous Hall conductivity can be used as an 
observable for spontaneous valley polarization. 
The Hall conductivity (per spin) is calculated by integrating $\Omega_{n, {\bm k}} $ 
over occupied quasiparticle states near the Dirac points,
\ber\label{hallcond}
\sigma_{\rm H} =  \frac{e^2}{\hbar}  \int \frac{d^{2} {\bm k}}{\left(2 \pi \right)^2} \sum_n f_{n,{\bm k}} \, \Omega_{n, {\bm k}} ~,
\eer
where $f_{n, {\bm k}}$ is the Fermi-Dirac distribution and a sum is carried over the 
band index $n$.

Fig.~\ref{fig:four} plots the Hall conductivities evaluated for this model and  
demonstrate that spontaneous valley polarization occurs for 
carrier densities smaller than $\sim 25 \times 10^{11}~{\rm cm}^{-2}$, 
and that nematic order occurs for carrier densities smaller than 
$\sim 9  \times 10^{11}~{\rm cm}^{-2}$ and again 
near the onset of minority valley occupation.  
The associated Berry curvatures and the orbital moments of the gapped chiral band edges
are represented together with the Hall conductivity to represent the $\bm k$ space resolved contribution.

\section{Optical conductivity signatures}
The presence of either valley polarization or nematic order should be observable via  
interband optical conductivity 
measurements. Circular dichroism measurements can detect valley polarization
because the associated optical transition matrix elements \cite{xiao} are valley dependent.
Valley-dependent population of states near the band edges then leads to 
valley-dependent Pauli blocking and contrast between the absorption of left and 
right circularly polarized light.  

Broken rotational symmetry in a nematic phase also has observable 
signatures in optical conductivity measurements comparing absorption 
of light that is is linearly polarized along different directions. 
For example, the real part of the conductivity per valley and spin for 
light polarized along the $x$ axis is given by 
\begin{equation}
\sigma_{xx}(\omega) = \frac{  \pi e^2}{\hbar \omega A} \sum_{{\bm k} }   {  \left|  W_{ {\bm k} \pm}  \right|^2}
 \left( f_{+, {\bm k}} - f_{-, {\bm k}} \right)   \, \delta (\hbar \omega - \Delta E_{\bm k})  
\end{equation}
where $W_{{\bm k} \pm } = \left<+ \, {\bm k}  \right|  j_x  \left| - \, {\bm k} \right> $, 
$\left| + {\bm k} \right>$ and $\left| - {\bm k} \right>$ are conduction and valence band quasiparticle states, 
$\Delta E_{\bm k}$ is the band splitting and $j_x = \partial H_{\bm k} / \partial {k_x}$ is the current operator in the $x$-direction.
At the band extrema, the quasiparticle states are symmetric and antisymmetric
combinations of the conduction bands of the low-potential graphene sheet
and the valence bands of the high-potential graphene sheet.  It follows that for states 
near the band edges $W_{{\bm k} \pm} \simeq \cos \theta_{\bm k}$.  
To illustrate how the momentum-direction dependence of the optical conductivity 
is sensitive to the spontaneous anisotropy of the nematic ground state, we evaluate  
the real part of the conductivity for incident light polarized along direction
$\varphi$ and a nematic state with conduction bands occupied between orientation angles 
$\theta_{\rm i}$ and $\theta_{\rm f}$.  We find that because absorption is reduced by Pauli-blocking 
the conductivity at the absorption edge is proportional to 
\begin{eqnarray}
\sigma_{\varphi}    \propto \left(     \theta_{\rm d}   
+ \sin   \theta_{\rm d}       \,        
\cos \left(  \frac{ \theta_{\rm s} }{2} +  \varphi \right)  \right)^2    
\end{eqnarray}
where $\theta_{\rm d} = \theta_{\rm f} - \theta_{\rm i}$ and $\theta_{\rm s} = \theta_{\rm f} + \theta_{\rm i}$.
Note that in the absence of nematicity ($\theta_{\rm d} = 2 \pi$) 
the optical absorption is independent of $\varphi$.  

\section{Discussion}  

Because the Fermi surface instabilities discussed in this paper
appear only at low carrier densities, comparable to or smaller than typical
disorder-induced density-fluctuation scales for bilayer samples on silicon oxide substrates,
and because large electric fields are favorable for their occurrence, 
we anticipate that momentum space condensation is at present a 
realistic possibility only for  
dual-gated bilayer graphene samples on h-BN substrates.~\cite{Dean}
We expect that trigonal warping of the unbalanced bilayer conduction bands 
will favor momentum space condensation over competing~\cite{CDW} density-wave 
instabilities.  Because the $\ell=1$ Pomeranchuk instability is 
likely only within states in which spin or valley polarization, or both, 
has already occurred, its appearance should be signaled most 
clearly by observables which detect reduced orientational symmetry,
for example polarization direction dependence in interband 
optical absorption.

{\it Acknowledgments.}
Financial support was received from Welch Foundation grant TBF1473, 
from DOE Division of Materials Sciences and Engineering grant DE-FG03-02ER45958, and from the Italian Ministry of Education, University, and Research (MIUR) through the program ``FIRB - Futuro in Ricerca 2010" Grant No. RBFR10M5BT (``PLASMOGRAPH: plasmons and terahertz devices in graphene").
We gratefully acknowledge assistance and computer time offered by the Texas Advanced Computing Center.


\begin{thebibliography}{77}
\bibitem{hartreecaveat} 
	This expression ignores the Hartree energy and is therefore valid only for uniform electron density states.
\end{thebibliography}
\end{document}